\documentstyle[prd,aps,eqsecnum,preprint]{revtex}

\begin{document}




\title{Flow equations for quark-gluon interactions 
in light-front QCD}

\author{Elena Gubankova, Chueng-Ryong Ji, Stephen R. Cotanch}


\address{Department of Physics, North Carolina State University,
Raleigh, NC 27695-8202}


\def\Journal#1#2#3#4{{#1} {\bf #2}, #3 (#4)}
 
\def\NCA{\em Nuovo Cimento}
\def\NIM{\em Nucl. Instrum. Methods}
\def\NIMA{{\em Nucl. Instrum. Methods} A}
\def\NPB{{\em Nucl. Phys.} B}
\def\PLB{{\em Phys. Lett.}  B}
\def\PRL{\em Phys. Rev. Lett.}
\def\PRD{{\em Phys. Rev.} D}
\def\PRB{{\em Phys. Rev.} B}
\def\PR{\em Phys. Rept.} 
\def\ZPC{{\em Z. Phys.} C}
\def\ZPB{{\em Z. Phys.} B} 
\def\APPB{{\em Acta Phys. Polon.} B}
\def\IJMPA{{\em Int.J.Mod.Phys.} A}
\def\AP{\em Ann. Physik}
\def\PTP{\em Prog. Teor. Phys.}
\def\JMP{\em J. Math. Phys.}

\maketitle

\begin{abstract}
The flow-equation method of continuous unitary transformations
is used to eliminate the minimal quark-gluon interaction 
in the light-front quantized QCD Hamiltonian.
The coupled differential equations in the two lowest Fock sectors  
correspond to the renormalization
of the light-front gluon mass and the generation 
of an effective quark-antiquark (as well as
gluon-gluon) interaction.   
The original gauge field coupling is completely
eliminated, even in the presence of degenerate states
connected by this interaction. 
Further, a more singular $1/q^4$ behavior
for the quark and gluon effective interactions
at small gluon momenta is obtained,
due to the asymptotic behavior of the effective
gluon mass for small cutoff. We discuss
the consequences of this asymptotic behavior
and possible confinement implications.

\end{abstract}


\section{Introduction}

The perturbative aspects of Quantum Chromodynamics (QCD)
were understood many years ago with the 
convincing documentation that QCD is asymptotically free. 
However, the calculational techniques for nonperturbative QCD
are still under developement. 
Thus, quantitative analysis of low energy and momentum 
transfer phenomena remains difficult
even though the qualitative features 
are reasonably described due to chiral symmetry 
and the phenomenological success of the constituent
quark picture.
Nevertheless, it is widely believed that pure Yang-Mills theory,
without dynamical quarks, exhibits 
confinement represented by a linear potential 
between static color sources. Confinement may be attributed 
to mass generation from transmutation of dimensions
in QCD.
Adding dynamical quarks also breaks 
chiral symmetry spontaneously. 
An ultimate aim of nonperturbative QCD 
studies is to understand both confinement and chiral symmetry 
breaking and how the constituent quark picture arises
from fundamental QCD.

In the past few years there were several studies  
addressing the issue of confinement and mass gap generation 
in the framework of the Schr{\"o}dinger picture
\cite{DiakonovZarembo,Karabali}. 
Using path integral techniques, ref.  
\cite{DiakonovZarembo} utilized
a vacuum wave functional ansatz suggested by Kogan and Kovner,
and integrated over all possible gauge configurations.
Since calculations are quite formidable,
this study only solves a field theory problem in $1+1$ dimensions 
\cite{Karabali} and restricts
$3+1$ dimension \cite{DiakonovZarembo} analyses to ground states.
For earlier investigations, see ref. \cite{Schr"odinger}.  

Alternatively, nonperturbative studies have used  
a Hamiltonian framework with the QCD Hamiltonian quantized 
in a specific gauge.
In particular the Coulomb gauge has recently 
been investigated \cite{Robertson,Coulomb}.
In this paper we also use a Hamiltonian approach but utilize
the light-front gauge, $A^+=0$ \cite{BrodskyPauliPinsky}. 
There are arguements that the light-front gauge may be the most 
suitable framework to study nonperturbative QCD 
\cite{BrodskyPauliPinsky}.
This conjecture is also supported by the success of phenomenological
constituent light-front quark models \cite{ChoiJi}.
To provide further insight concerning this issue,
we have applied the flow equations to the light-front
QCD Hamiltonian to generate dynamical gluons
and quarks as well as their effective interactions.

Previous light-front studies of confinement and bound states 
have been conducted using the methods 
of similarity renormalization \cite{BrisudovaPerry}
and transverse lattice \cite{BvdSandeDalley}. 
Significantly, light-front QCD in $3+1$ 
rigorously contains a confining interaction in the form of the instantaneous 
four-fermion interaction, $1/q^{+~2}$, 
which is the complete confining interaction in $1+1$ QCD
for the light-front spatial dimension, $x^-$ . 
Wilson suggested that a starting point for analyzing full QCD 
(with confinement) in light-front coordinates is the 
light-front infrared divergences. 
When an infrared (IR) cutoff is introduced,  
appropriate counterterms are necessary to restore 
the attending physics below the cutoff.
Based on light-front power counting, 
these counterterms can involve color charge densities 
and exhibit an unknown 
nonlocal behavior in the transverse direction which represents 
a possible source
for transverse confinement. 
In the similarity renormalization approach, it has been claimed
\cite{BrisudovaPerry}
that IR divergences from the instantaneous gluon exchange 
potential are not completely cancelled leading to a remnant potential  
that increases logarithmically with either increasing
separations $|x^-|$ or $|x_{\perp}|$.
However, the issue of a nonzero gluon mass and local gauge 
invariance is not yet completely understood 
and the task of restoring rotational symmetry is still difficult
to achieve.

In this work, we also introduce an IR 
longitudinal cutoff and generate a light-front counterterm
which sets a scale for a dynamical mass gap
and string tension. While our calculations for 
the flow equation are only to order $g^2$,  
our results challenge the conventional notion that 
weak-coupling Hamiltonians derived from QCD have only
Coulomb-like potentials, and definitely no 
confining interactions.
Introducing a longitudinal IR cutoff in light-front dynamics
makes it impossible to create particles from a bare vacuum 
by a translationally invariant Hamiltonian and thus
the number of constituents in a given eigenstate is fixed. 
However, light-front counterterms for the longitudinal IR cutoff
dependence can generate a nonzero amplitude of particle creation and are
therefore a possible source for features 
associated with a nontrivial vacuum structure
in equal-time dynamics, including confinement and 
spontaneous symmetry breaking. Note that small light-front 
longitudinal momentum fraction $x$
corresponds to high light-front energy. Therefore, in order
to remove small $x$ divergences and maintain the cutoff-independence, 
one should use a renormalization group,
which is appropriate to high energies.
 
In this work we address the above issue using the
flow equations method \cite{Wegner}. 
More specifically, we adopt Susskind's idea of the 
'long arm' of the vacuum \cite{SusskindBurkardt} 
which addresses chiral symmetry breaking 
in the light-front formulation.
In the parton model, one pictures a high momentum hadron 
as a collection of constituents
each also having large momentum,
such that under a boost, both the hadron and the partons
change their momenta by the same factor. 
Therefore, one can formulate an effective field theory on the axis
of the light-front longitudinal momentum fraction $x$ (or on the  
axis of rapidity which is logarithm of $x$). The partons, both valence and 
sea, have positive $x$ and, according to Feynman and Bjorken, 
are distributed along the $x$-axis 
with the density of sea partons increasing for small 
$x$ according to $dx/x$. Here, the vacuum is
at $x=0$. The fundamental property of light-front Hamiltonians 
is that under a rescaling of the light-front momentum,
$x\rightarrow \alpha x$, the light-front Hamiltonian scales 
$H\rightarrow H/\alpha$. 
This may be interpreted as a dilatation symmetry
along the $x$-axis.  
This symmetry only holds classically and is broken
on the quantum level by an anomaly.
Now, the long arm of the vacuum occurs because the coupling,{\em i.e.}
the interaction between neighboring partons, gets stronger and stronger
as one approaches $x=0$ so rapidly that the system is able to hold
itself together despite the fact that there is an infinite number of
steps between $x=0$ and finite $x$\cite{SusskindBurkardt}.
To illustrate the long arm effect of the vacuum,
we introduce a cut-off for small $x$.
A natural cutoff is provided by $\delta x=\varepsilon^{+} x$, 
the minimal spacing between constituents,
which plays the role of UV-regulator. 
As long as the density of partons 
on the rapidity axis is not infinite, 
$\varepsilon^+$ or $\delta x$ is finite
and one obtains finite matrix elements.
The cutoff $\delta x$ breaks 
dilatation symmetry in the $x$-axis
and generates an energy scale,
or mass gap governing the strength
of the effective interaction between quarks,
in our case a string tension in the quark-antiquark potential.
Hence the long arm of the 'light-front vacuum' enables
formation of a $q\bar{q}$ bound state through breaking 
of an internal symmetry, analogous 
to the creation of Cooper pairs in a superconducter.
Because results must be independent of the cutoff,
renormalization group equation is required, which in this work
is provided by the Hamiltonian flow equations method 
\cite{Wegner}.
  
Incorporating effects from small $x$ 
into an effective light-front Hamiltonian is equivalent to integrating
over the high light-front-energy modes in the asymptotically free domain. 
In terms of the renormalization group, regulating small $x$ introduces 
a mass gap, which together with asymptotic freedom leads
to a renormalization group invariant scale and 
dimensional transmutation along the $x$-axis. 
This transmutation in turn produces 
a linear effective interaction more singular than
the Coulomb potential.

At the technical level our thrust is to use the flow equations
to renormalize the gluon energy and eliminate
the coupling between the gluon and quark Fock sectors.
We focus upon zero gluon momentum and obtain
a gap equation for the renormalized light-front gluon mass
$\mu(\lambda)$. This equation can be solved by imposing 
a renormalization condition, $\mu_{ren}(\lambda=\lambda_0)=\mu_0$,
where $\mu_0$ is the 'physical' value.
The renormalized
effective gluon mass, $\mu_{ren}(\lambda)$, 
is obtained by introducing a mass counterterm.
As a result, the asymptotic behavior of 
$\mu_{ren}(\lambda)$ 
is obtained for small cutoff $\lambda$ which
approaches the renormalization point $\lambda_0$ 
from above ($\lambda\geq\lambda_0$).
The properties of $\mu_{ren}(\lambda)$
ensure the quark-gluon coupling is 
eliminated even in the degenerate case
of zero gluon momenta. A similar idea was used 
for the spin-boson model
by Kehrein, Mielke and Neu \cite{KehreinMielkeNeu}, 
who argued that the coupling to a bosonic bath
is always eliminated by renormalizing the tunneling
frequency. Also, in complete analogy 
to our problem, 
Lenz and Wegner \cite{LenzWegner} showed
for interacting electrons in BCS-theory,  
that the elimination of electron-phonon coupling 
for all states is a direct consequence 
of renormalizing the phonon frequency. 
Finally, the flow of the gluon mass with $\lambda$ produces
for small momentum $q$ a potential of form $1/q^4$
which is more singular than the Coulomb $1/q^2$. 

In the next section (Section II),  
we consider flow equations as a renormalization 
group transformation for Hamiltonians and formulate them for one 
and two-body sectors explicitly.
In Section III
the flow equations are applied to the light-front QCD Hamiltonian 
to generate a gluon gap equation and an effective $q\bar{q}$
(as well as $gg$) interaction. This section also addresses solving
these equations. Concluding discussion follows in Section IV.

\section{Flow equations for gauge theories}

\subsection{Flow equations as renormalization group transformations}

The basic element of the renormalization group transformation 
is a unitary transformation that renders the Hamiltonian
matrix band-diagonal; i.e. matrix elements with energy differences 
$|E_i-E_j|$ exceeding a cutoff
$\lambda$ are eliminated \cite{GlazekWilson}. 
This procedure converges well when there is
a hierarchy of scales in the problem. The goal is to decouple 
the high- and low-energy scales of the band-diagonal effective Hamiltonian, 
which is renormalized order by order in perturbation theory.
Using $N$th order perturbation theory to connect
high and low energy states of an effective renormalized Hamiltonian  
with ultraviolet regulating cutoff $\Lambda$ (which is the size
of the full Hamiltonian matrix in the energy space, eventually 
taken to infinity), requires $N=2(\Lambda-\lambda)/\lambda$.
Assuming a coupling constant $g<1$,
$g^{2(\Lambda-\lambda)/\lambda}\ll 1$ 
for $\Lambda\rightarrow\infty$. 
One has therefore isolated the low energy scale effective Hamiltonian
which can then be diagonalized nonperturbatively
for the few lowest eigenstates. This should be 
simpler than solving the full Hamiltonian matrix exactly.
The unitary transformation which connects Hamiltonian matrices 
with different widths, 
$H(\lambda_2,\Lambda)= U(\lambda_1,\lambda_2;\Lambda)
H(\lambda_1;\Lambda)U^{\dagger}(\lambda_1,\lambda_2;\Lambda)$,
is the renormalization group transformation
formulated by Glazek and Wilson \cite{GlazekWilson} 
called the similarity renormalization.

In terms of an effective Hamiltonian, one removes large energy differences 
by squeezing the width of the band, $\lambda_2< \lambda_1\ll\Lambda$. 
This is done sequencially from high to low energy to
construct a low-energy effective Hamiltonian.
Since the light-front energy has an inverse relationship 
to the light-front momentum, $E_{LF}\sim 1/P^+$, 
this scaling to lower $\lambda$
is equivalent to moving from small $x$ (high energy) to large $x$ 
(low energy) for an effective light-front theory 
formulated on the $x$-axis.
Elimination of large energy differences is governed by 
a dimensionless formfactor,
called the similarity function, $f(|E_i-E_j|/\lambda)$, 
which is of order unity for small arguments (no elimination),
and approaches zero for large arguments (complete elimination
of the off-diagonal matrix elements). Note that squeezing 
the band width to zero
is equivalent to exactly diagonalizing the Hamiltonian matrix. 

The flow equation approach entails an
infinitesimal unitary transformation written in
differential form
\begin{eqnarray}
\frac{dH(l)}{dl}=[\eta(l),H(l)]
\nonumber\\
\eta(l)=[H_d(l),H_r(l)]
\,,\label{eq:1.1}\end{eqnarray}
where $l$ is the flow parameter related to the cutoff scale $\lambda$
by $l=1/\lambda^2$. 
The choice for the transformation generator $\eta$   
has been suggested by Wegner \cite{Wegner} as the commutator
of the diagonal, $H_d$, and off-diagonal (rest), $H_r$, 
Hamiltonian components. As shown by Wegner,
for $l\rightarrow \infty$ the off-diagonal part
of Hamiltonian is eliminated, producing 
an effective diagonal Hamiltonian. 

One of the purposes for renormalization 
is to remove the UV divergent
high-energy contributions. Since they arise from 
the particle-number-changing terms of the Hamiltonian 
(at least in gauge theories), 
only these terms should be eliminated.
Wegner suggested assigning the off-diagonal part to 
the particle-number-changing term $H_r=H_{n\rightarrow m}$, 
and the diagonal part to the 
particle-number-conserving term $H_d=H_{n\rightarrow n}$. 
Thus instead of diagonalization, one can 
implement flow equations to block-diagonalize
the Hamiltonian in particle number space. 
Also, block-diagonalization in particle number space 
precludes convergence problems associated with
exact diagonalization in energy space \cite{Wegner}.
Most noteworthy is the
practical aspect of block-diagonalization 
when applied to field theory.
By block-diagonalizing the Hamiltonian in particle number space
the high and low Fock sectors are uncoupled,
enabling separate eigenstate problems with an effective Hamiltonian 
in each particle number sector. For most calculations,
solving an effective sector-uncoupled Hamiltonian 
should be simpler than solving the original Hamiltonian
\cite{GubankovaWegner}.
Note, however, that the flow equations eliminate 
the particle number changing matrix elements not in one step,
but rather continuously for different energy differences,
sequencially from high to low energy
(matrix elements between degenerate states, $E_i=E_j$,    
can also be eliminated by flow equations 
as discussed in the next subsection).
Here, the link between the similarity and flow equation schemes
and renormalization clearly emerges.
The distinguishing feature for 
flow equations, however, is the separation of 
the particle-number-conserving and
particle-number-changing terms, with only 
the particle-number-changing terms
contributing in the renormalization of an effective Hamiltonian.

In this work, we run flow equations for the two lowest Fock sectors
($|q\bar{q}\rangle$ and $|g\rangle$) 
in a gauge theory
and obtain equations for the single gluon energy and $q\bar{q}$ 
effective interaction.

\subsection{Flow equations in the two lowest Fock sectors}

We consider a canonical Hamiltonian operator for a gauge field theory,
where the gauge field is minimally coupled to the matter field
(for example, minimally coupled QED or simple abelian QCD). 
In terms of bare quark and gluon fields 
the eigenfunctions of this Hamiltonian
contain infinitely many Fock states. For
the two lowest Fock states (neglecting the rest),
the Hamiltonian matrix in particle number space is 
\begin{eqnarray}
H=\left(
\begin{array}{cc}
PHP & PHQ \\
QHP & QHQ
\end{array} \right)
\,,\label{eq:1.2}\end{eqnarray}
where $P$ and $Q$ are projection operators for one- and two-body
Fock states. For abelian, minimally coupled QCD, $P$ projects 
on a one-gluon state and $Q$ on a quark-antiquark pair,
$P|\psi\rangle=|g\rangle$ and $Q|\psi\rangle=|q\bar{q}\rangle$.
One-quark states are omitted because there are no dynamical quarks 
in this analysis (we prefer to disentangle complexities
from chiral symmetry breaking). Matrix elements of the operator
$PHQ$ describe quark-gluon minimal coupling, the $PHP$ term is 
a gluon effective energy and $QHQ$ describes the $q\bar{q}$
effective interaction. Flow equations for particle number changing
matrix elements, $h_{pq}$, are given by
\begin{eqnarray}  
\frac{dh_{pq}(l)}{dl} &=& -\left(
E_p(l)-E_q(l) \right) \eta_{pq}(l)
\nonumber\\
\eta_{pq}(l) &=& -\frac{h_{pq}(l)}{E_p(l)-E_q(l)}\frac{d}{dl}
\left( \ln f(z_{pq}(l)) \right)
\nonumber\\
z_{pq}(l) &=& l\left( E_p(l)-E_q(l) \right)^2
\,,\label{eq:1.3}\end{eqnarray}
where $p$ and $q$ span all (free single particle) energy 
states in the $P$ and $Q$ subspaces, respectively; 
$E_p(l)$ and $E_q(l)$ are diagonal matrix elements 
of block Hamiltonians $PHP$ and $QHQ$. In the first equation
the off-diagonal matrix elements $h_{pp'}$ and $h_{qq'}$
are neglected to leading order in coupling $O(h_{pq})$
(this uncouples the flow equations for $H_d$ and $H_r$ terms).
With the generator $\eta_{pq}$ defined by Eq.~(\ref{eq:1.3}), 
elimination of the coupling between the $P$ and $Q$ sectors  
\begin{eqnarray}
h_{pq}(l)=h_{pq}(0)f(z_{pq}(l))
\,,\label{eq:1.4}\end{eqnarray} 
is governed by the similarity function $f(l(E_p(l)-E_q(l))^2)$, 
which vanishes for matrix elements with energy differences 
exceeding the cutoff $\lambda$, $|E_p(l)-E_q(l)|\gg 1/\sqrt{l}=\lambda$.
Also, $h_{pq}(0)$ is the initial value, and $E_p(l)$ and $E_q(l)$ 
flow too (see Eq.~(\ref{eq:1.6})). 
Eliminating the coupling $h_{pq}$
generates effective interactions in the particle number conserving
$P$ and $Q$ sectors. The corresponding flow equation reads
\begin{eqnarray}
\frac{dh_{pp'}(l)}{dl}=\sum_q\left(
\eta_{pq}(l)h_{qp'}(l) -h_{pq}(l)\eta_{qp'}(l)
\right)
\,,\label{eq:1.5}\end{eqnarray}
with a similar equation for $h_{qq'}$.    
Using the generator defined in Eq.~(\ref{eq:1.3})
these equations reduce to 
\begin{eqnarray}
\frac{dE_p(l)}{dl} &=& -\sum_{q}\frac{1}{E_p(l)-E_q(l)}
\frac{d}{dl}\left( h_{pq}(l)h_{qp}(l) \right)
 \label{eq:1.6} \\
\frac{dh_{qq'}(l)}{dl} &=& -\sum_{p}\left(
\frac{dh_{pq}(l)}{dl}\frac{1}{E_q(l)-E_p(l)}h_{pq'}(l) +
h_{qp}(l)\frac{1}{E_{q'}(l)-E_p(l)}\frac{dh_{pq'}(l)}{dl}
\right) \nonumber
\,,\end{eqnarray}
with the coupling $h_{pq}$ given by Eq.~(\ref{eq:1.4}). Note,  
for the first equation in Eq.~(\ref{eq:1.6}) 
only the diagonal matrix elements in the $P$ space enter.
In our application these equations represent
the gap equation for an effective gluon energy
and an equation for an effective $q\bar{q}$-interaction,
both investigated in the next section. 

Generally, the set of Hamiltonian flow equations is not finite 
and the equations are coupled by kernels, which 
also flow with $l$, that are only known after solving the flow equations.  
Obviously, practical computations require truncating 
to the few lowest sectors, assuming that higher sectors 
are not very important at low energies. 
In the light-front framework, this approximation is valid
because pair creation from the light-front vacuum is forbidden
and generally the higher Fock components (with large particle number)
carry large light-front energies. Neglecting the high Fock components
reduces the problem to an effective low-energy theory. 
In this way, we close the set of equations 
and reduce the number of unknown couplings, leaving only
the canonical operator couplings. 
Note that
the truncation in number of particles participating in intermediate
states is not equivalent to perturbation theory 
in coupling constant, but rather is closer to the
Tamm-Dancoff approach. The flow equations are still  
coupled, and should be solved selfconsistently
including the flow dependences of the couplings.

In this work on minimally coupled gauge theory, 
the set of flow equations was truncated at the two-body sector,
including at most three-particles in intermediate states.
Hence the renormalization of the QCD strong coupling constant 
is not considered, but the single particle energy, 
Eq.~(\ref{eq:1.6}) for $E_p(l)$, is renormalized. 
In particular, as shown in the next 
section, the renormalization of a gluon energy is important for  
constructing an effective $q\bar{q}$-interaction 
over all energy ranges.

Flow equations not only eliminate the quark-gluon coupling,
that renders the Hamiltonian block-diagonal, they also
permit renormalization. Renormalizing the gluon effective energy
yielding $E_p(l)$ leads to two important consequences. 
First, the quark-gluon coupling is eliminated for all energies, 
even for degenerate states.
Second, eliminating the quark-gluon coupling with $E_p(l)$
for degenerate states, the effective $q\bar{q}$ interaction
is obtained of the form $1/q^4$, which is more singular than 
the perturbative result, $1/q^2$.
This potential provides confinement and bound state of quarks.   

We conclude this section with a synopsis of previous flow equation
applications in other fields. The situation with degenerate states 
was first treated by flow equations in solid state physics. 
Lenz and Wegner showed for systems of interacting electrons 
\cite{LenzWegner}, that the original electron-phonon coupling 
can be completely eliminated, even when the states connected 
by this interaction are degenerate. 
Lenz and Wegner found \cite{LenzWegner} 
that the induced electron-electron interaction differs 
from Fr{\"o}hlich's, whose unitary transformation is based 
on second order perturbation theory. Moreover, this interaction 
is attractive for any momentum, binding
electrons in Cooper pairs.
Also, Kehrein, Mielke and Neu \cite{KehreinMielkeNeu}
have shown for the spin-boson problem, that flow equations
allow a complete elimination of the coupling to the bosonic bath 
even for real processes. Finally, Kehrein and Mielke obtained
similar modifications due to an $l$-dependent generator 
by eliminating the hybridization term in the Anderson model 
\cite{KehreinMielke}.
The authors showed that their approach generates a spin-spin
interaction which differs from the one obtained by the famous
Schrieffer-Wolff transformation. Their induced interaction
has the right high-energy cutoff, as compared to 
Schrieffer-Wolff's result. Thus, within flow-equation approach 
it is possible to obtain new results which can not be obtained 
by perturbation theory.

In the next section, we use the above formulation to solve
the flow equations for the effective gluon energy and 
the quark-antiquark interaction (see Eq.~(\ref{eq:1.6})) using 
the light-front quantized QCD Hamiltonian.

\section{Flow equations in light-front QCD}

\subsection{Gluon gap equation}

In the light-front formulation, the flow equation 
for a single particle energy $p^-=(p_{\perp}^2+\mu^2(\lambda))/p^+$
is actually written for the mass $\mu^2(\lambda)$ since 
the term $p_{\perp}^2/p^+$ is independent of flow. 
The set of the coupled light-front equations for 
the cutoff dependent quark and gluon masses 
was first derived by Glazek \cite{Glazek}. We uncouple 
this set of equations by assuming that the quark mass
does not flow with the cutoff, $m(\lambda)=m$,
where $m$ is the current quark mass. 
The light-front gluon gap equation is
\begin{eqnarray}
\frac{d\mu^2(\lambda)}{d\lambda} =
&-& 2T_fN_f\int_0^1\frac{dx}{x(1-x)}
\int_0^{\infty}\frac{d^2k_{\perp}}{16\pi^3}
g_q^2(\lambda)\frac{1}{Q^2_2(\lambda)}
\frac{df^2(Q^2_2(\lambda);\lambda)}{d\lambda}
\nonumber\\
&\times&
\left(
\frac{k_{\perp}^2+m^2}{x(1-x)}
-2k_{\perp}^2\right)
\nonumber\\
&-& 2C_a\int_0^1\frac{dx}{x(1-x)}
\int_0^{\infty}\frac{d^2k_{\perp}}{16\pi^3}
g_g^2(\lambda)\frac{1}{Q_1^2(\lambda)}
\frac{df^2(Q_1^2(\lambda);\lambda)}{d\lambda}
\nonumber\\
&\times&
\left(
k_{\perp}^2(1+\frac{1}{x^2}+\frac{1}{(1-x)^2})
\right)
\,,\label{eq:2.1}\end{eqnarray}
where
\begin{eqnarray}
Q_1^2(\lambda) &=& \frac{k_{\perp}^2+\mu^2(\lambda)}{x(1-x)}
-\mu^2(\lambda)
~,~
Q_2^2(\lambda) =
\frac{k_{\perp}^2+m^2}{x(1-x)}
-\mu^2(\lambda)
\,.\label{eq:2.2}\end{eqnarray}
Here, the group factors for $SU(N_c)$ are
$T_f\delta_{ab}={\rm Tr}(T^aT^b)=\frac{1}{2}\delta_{ab}$
and the number of flavors is $N_f=6$.
The adjoint Casimir is defined by 
$C_a\delta_{ab}=f^{acd}f^{bcd}=N_c\delta_{ab}$
with the number of colors $N_c=3$
(the subscripts in $C_a$ and $N_c$ should not be confused 
with the group indices $a$ and $c$).
In the integral kernel, the gluon couples to quark-antiquark pairs
and gluon pairs with quark-gluon coupling, $g_q(\lambda)$,
and three-gluon coupling, $g_g(\lambda)$, respectively.
For non-zero $\lambda$ these couplings are different 
functions of momenta.  
The light-front momentum flowing in the loops 
has components $(x,k_{\perp})$.
In our derivation, 
the connection between flow parameter and the cutoff,
$l=1/\lambda^2$, is used. 

In Eq.~(\ref{eq:2.1}), the effective gluon mass is defined 
at transverse gluon momentum
$p_{\perp}=0$, as in light-front perturbation theory 
for the gluon mass correction 
\cite{WilsonWalhoutHarindranathZhangPerryGlazek}. 
Following other gluon gap equation studies, 
we assume the effective gluon mass vanishes for large
gluon momenta. Therefore, even though the effective gluon mass 
depends on momentum, only its $\lambda$ cutoff dependence 
at zero momentum is given by Eq.~(\ref{eq:2.1}), i.e.
$\mu(\lambda,p(p_z,p_{\perp})=0)$. 
However, the $\lambda$ dependence of $\mu$ is 
the only relevant renormalization issue.

Generally, it is difficult to solve Eq.~(\ref{eq:2.1})
because the running couplings, $g_q(\lambda)$ and $g_g(\lambda)$,
are not known. Therefore, the coupling cutoff dependence is neglected.
Also, the initial condition for Eq.~(\ref{eq:2.1}) is not known.
Accordingly, the following two renormalization conditions 
are imposed to determine the running gluon mass
$\mu(\lambda)$.  
First, the effective Hamiltonian at scale $\lambda$ 
has eigenstates with eigenvalues  
$p^-=(p_{\perp}^2+\mu_0^2)/p^+$ \cite{Glazek}, satisfying 
\begin{eqnarray}
\frac{p_{\perp}^2+\mu_0^2}{p^+}\langle p'|p\rangle &=&
\frac{p_{\perp}^2+\mu^2(\lambda)}{p^+}\langle p'|p\rangle
\nonumber\\
&-& \int^{\lambda} d\lambda' \sum_{q}
\left(\eta_{p'q}(\lambda')h_{qp}(\lambda')
-h_{p'q}(\lambda')\eta_{qp}(\lambda')\right) 
\,,\label{eq:2.4}\end{eqnarray}
where $\mu_0$ denotes the 'physical' gluon mass,
and $|p\rangle$ denotes a single effective gluon state 
($P$ subspace, Eq.~(\ref{eq:1.2})) 
with momentum $(p^+,p_{\perp}$) and 
$\langle p'|p\rangle =16\pi^3p^+\delta^{(3)}(p'-p)$.
The generator $\eta_{pq}$, given by Eq.~(\ref{eq:1.3}), 
eliminates the quark-gluon (three-gluon) coupling $h_{pq}$.
Second, the effective gluon mass, renormalized in second order 
perturbation theory, equals the 'physical' mass
\begin{eqnarray}
\mu_{ren}^2(\lambda=\lambda_0)=\mu_0^2
\,,\label{eq:2.5}\end{eqnarray}
at the renormalization point $\lambda_0$.      

From the definition of flow equations, a resulting effective 
Hamiltonian is given in the limit of 
the flow parameter $l\rightarrow\infty$,
or $\lambda=0$. 
In the parton picture described in the introduction,
the boost may be regarded as a renormalization
group operation and the renormalization group fixed point
can be identified as the infinite momentum limit.
The corresponding renormalization group fixed point
in the flow equations is then $\lambda=0$.
Therefore, in Eq.~(\ref{eq:2.5}), the renormalization point 
is set to zero, $\lambda_0=0$, at the end of calculations.
Also, the 'physical' gluon mass, 
which is used as an arbitrary mass parameter, 
can be taken to zero to restore gauge invariance (see below). 

Even with these simplifications, solving Eq.~(\ref{eq:2.1}) 
is still quite involved.
The solution can only be found numerically. 
In order to proceed analytically,
a mass parameter $\mu_0$
('physical' mass), that corresponds to the choice of 
the renormalization point 
at $p^2=\mu_0^2$,
is substituted for $\mu(\lambda)$ 
in the integral
kernel on the right-hand-side of Eq.~(\ref{eq:2.1}). 
The same procedure is used to calculate 
the perturbative mass correction in \cite{ZhangHarindranath}.
Therefore the energy differences in Eq.~(\ref{eq:2.2}) 
are given by  
\begin{eqnarray}
\tilde{Q}_1^2=\frac{k_{\perp}^2+\mu_0^2}{x(1-x)}-\mu_0^2
~,~
\tilde{Q}_2^2=\frac{k_{\perp}^2+m^2}{x(1-x)}-\mu_0^2
\,.\label{eq:2.6}\end{eqnarray}
Using Eq.~(\ref{eq:2.6}),
Eq.~(\ref{eq:2.1}) is solved iteratively and 
$\mu^2(\lambda)=\mu_0^2$ is the first iteration.

Another problem in Eq.~(\ref{eq:2.1}) is 
that the loop integrals have an UV divergence in the 
transverse directions $k_{\perp}$ and an IR divergence 
in the longitudinal direction $x$. 
The flow equations naturally regulate these divergences 
via the similarity function in each vertex 
(for example, the gluon loop
in Eq.~(\ref{eq:2.1}) is regulated by 
$Q_1^2(\lambda)\leq \lambda^2$). 
This type of regularization is known as 'Jacobi cutoffs', 
because of the Jacobi momenta of a constituent
(it is also called 'global' regularization in 
\cite{ZhangHarindranath}).
The advantage of this regularization in the light-front approach
is that it preserves both transverse 
and longitudinal boost invariance
\cite{WilsonWalhoutHarindranathZhangPerryGlazek}. 
For a nonzero mass
such as a current quark mass in the quark loop, 
this regularization
restricts $k_{\perp}$ and the light-front $x$ integrations.
Thus, the similarity function $f(\tilde{Q}_2^2;\lambda)$, 
with $\tilde{Q}_2^2$ given by Eq.~(\ref{eq:2.6}), 
ensures that
$0\leq k_{\perp}\leq (\lambda^2+\mu_0^2)x(1-x)-m^2$
and $m^2/(\lambda^2+\mu_0^2)\leq x\leq 
1-m^2/(\lambda^2+\mu_0^2)$. However, for zero gluon mass
in the gluon loop, $k_{\perp}$ is restricted from above and $x$
runs the entire range $0\leq x\leq 1$, 
contributing divergent terms at $x=0$ and $x=1$
(see the similarity function $f(\tilde{Q}_1^2;\lambda)$).
Even when the instantaneous diagrams are included, 
the gluon loop is still divergent in $x$
\cite{ZhangHarindranath}. For a massive case 
the cutoff $\lambda$ 
has a lower bound from a mass in the theory,
$\lambda\geq m\neq 0$, that limits the $x$-integration. 
The reason why flow equations
do not regulate the light-front divergences 
in a massless case is because the cutoff $\lambda$ 
can evolve to $\lambda=0$ with the similarity function 
$f(z\rightarrow\infty)=1$, and $x$ spans the entire 
$0$ to $1$ range, 
where the loop integral with massless intermediate states 
diverges at $x=0$ and $x=1$.
To regulate these divergences 
in a massless case, Zhang and Harindranath suggested restricting the 
$k_{\perp}$ integration also from below by some scale $u$.
This is equivalent in our approach to integrating the flow equation, 
Eq.~(\ref{eq:2.1}),
for finite limits, from $\lambda$ to $u$  
\cite{ZhangHarindranath}, which restricts
intervals in the $x$-axis to above zero 
and below one. It was also shown in 
\cite{WilsonWalhoutHarindranathZhangPerryGlazek}
and \cite{ZhangHarindranath} that for the scale $u$,
even if $\mu_0=0$ is zero, the gluon mass correction does not vanish.  
Effectively, introducing the scale $u$ mimics the situation 
of a non-zero mass, $m=u$, in intermediate states. 

In terms of effective light-front theory formulated on the $x$-axis,
the Hamiltonian below the first light-front cutoff, 
$H_{0\leq x\leq\varepsilon}$, as well as the Hamiltonian
above the second cutoff $x=1$, 
$H_{(1-\varepsilon)\leq x\leq 1}$, describe 
the strongly correlated high energy states. 
Thus, they can be replaced by the Hamiltonian vacuum expectation
value, since strongly coupled configurations are frozen.
This v.e.v. replacement in gluodynamics
is consistent with a composite field, $\phi$, 
creating $0^+$ glueballs having finite v.e.v. \cite{Karabali}.
The Hamiltonian in the intermediate region, 
$H_{\varepsilon\leq x\leq (1-\varepsilon)}$,
is then treated by flow equations. 

Integrating the gluon flow equation, Eq.~(\ref{eq:2.1}), 
for finite limits $[u;\lambda]$,
yields
\begin{eqnarray}
\mu^2(\lambda) - \mu^2(u) =
&-& 2g^2T_fN_f\int_0^1dx\int_0^{\infty}
\frac{d^2k_{\perp}}{16\pi^3}
\left(f^2(\tilde{Q}_2^2;\lambda)-f^2(\tilde{Q}_2^2;u)\right)
\nonumber\\
&\times&
\left(
\frac{\mu_0^2(2x^2-2x+1)}{k_{\perp}^2+m^2-x(1-x)\mu_0^2}
+\frac{2m^2}{k_{\perp}^2+m^2-x(1-x)\mu_0^2}\right.
\nonumber\\
&+&\left. (-2+\frac{1}{x(1-x)}) \right)
\nonumber\\
&-& 2g^2C_a\int_0^1dx
\int_0^{\infty}\frac{d^2k_{\perp}}{16\pi^3}
\left(f^2(\tilde{Q}_1^2;\lambda)-f^2(\tilde{Q}_1^2;u)\right)
\nonumber\\
&\times&
\left(1+\frac{\mu_0^2 x(1-x)}
{k_{\perp}^2-x(1-x)\mu_0^2+\mu_0^2}
-\frac{\mu_0^2}{k_{\perp}^2-x(1-x)\mu_0^2
+\mu_0^2}\right)
\nonumber\\
&\times&
\left(1+\frac{1}{x^2}+\frac{1}{(1-x)^2}\right)
\,,\label{eq:2.7}\end{eqnarray}
where the renormalization point is $q^2=\mu_0^2$, and
the energy differences are defined in Eq.~(\ref{eq:2.6}).
Though the quark loop does not diverge with $x$,
the gluon flow equation is integrated the same way 
for the gluon and quark loops.
Similarity functions restrict the transverse momenta to  
$k_{\perp min}\leq k_{\perp}\leq k_{\perp max}$, 
with
\begin{eqnarray}
k_{\perp max} &=& (\lambda^2+\mu_0^2)x(1-x)-\mu_0^2
\nonumber\\
k_{\perp min} &=& (u^2+\mu_0^2)x(1-x)-\mu_0^2
\,,\label{eq:2.8}\end{eqnarray}
and for $x$  
\begin{eqnarray}
\frac{\mu_0^2}{u^2+\mu_0^2}\leq x\leq
1-\frac{\mu_0^2}{u^2+\mu_0^2}
\,.\label{eq:2.9}\end{eqnarray}
Analagous expressions hold for the quark loop. 
 
Integration over momenta $(x,k_{\perp})$
in Eq.~(\ref{eq:2.7}) yields
\begin{eqnarray}
\mu^2(\lambda)-\mu^2(u) =
&-& \frac{g^2T_fN_f}{4\pi^2}\left(
\frac{1}{3}
\mu_0^2\ln\frac{\lambda^2}{u^2}
+m^2\ln\frac{\lambda^2}{u^2}
+\frac{1}{3}(\lambda^2-u^2)
\right)
\label{eq:2.10} \\
&-& \frac{g^2C_a}{4\pi^2}\left(\mu_0^2\ln\frac{\lambda^2}{u^2}
(-\frac{u^2}{\mu_0^2}
+\ln\frac{u^2}{\mu_0^2}-\frac{5}{12})
+(\lambda^2-u^2)(\ln\frac{u^2}{\mu_0^2}-\frac{11}{12})
\right)
\nonumber
\,.\end{eqnarray}
Assuming the renormaliation conditions of 
Eqs.~(\ref{eq:2.4}) and (\ref{eq:2.5}), 
the effective gluon mass at scale $\lambda$ is
\begin{eqnarray}
\mu^2(\lambda) &=& 
\mu_0^2+\delta\mu_{PT}^2(\lambda)
+\delta\mu^2(\lambda,\lambda_0)
\,.\label{eq:2.11}\end{eqnarray}
Here, Eq.~(\ref{eq:2.4}) fixes the integration constant 
(when integrating the flow equation (\ref{eq:2.1})) to $\mu_0^2$. 
In Eq.~(\ref{eq:2.11}) the perturbative term
reproduces the result from light-front perturbation theory
\cite{ZhangHarindranath}, i.e.
\begin{eqnarray}
\delta\mu_{PT}^2(\lambda)=-\frac{g^2}{4\pi^2}
\lambda^2 \left(
C_a (\ln\frac{u^2}{\mu_0^2}-\frac{11}{12})
+T_fN_f\frac{1}{3}\right)
\,.\label{eq:2.12}\end{eqnarray}
Renormalizing the effective Hamiltonian 
to second oder $O(g^2)$,
the perturbative mass correction is absorbed by 
the mass counterterm,
$m_{CT}^2(\Lambda_{UV})=-\delta\mu_{PT}^2(\Lambda_{UV})$
with $\Lambda_{UV}\rightarrow\infty$,
and the renormalized effective gluon mass is 
\begin{eqnarray}
\mu^2_{ren}(\lambda)=\mu^2(\lambda)+m_{CT}^2(\lambda)
\,,\label{eq:2.13}\end{eqnarray}
for $\lambda=\Lambda_{UV}\rightarrow\infty$.
Though perturbative renormalization is applied at large
cutoff scales, $\Lambda_{UV}$, we assume that the leading cutoff
dependence in second order 
is absorbed by the mass counterterm for all $\lambda$,
and the renormalized gluon mass is given by Eq.~(\ref{eq:2.13}) 
for any $\lambda$. 
Explicitly from Eq.~(\ref{eq:2.10}), the renormalized gluon mass reads
\begin{eqnarray}
\mu_{ren}^2(\lambda) &=& \mu_0^2+\delta\mu^2(\lambda,\lambda_0)
= \mu_0^2+\sigma(\mu_0,u)\ln\frac{\lambda^2}{\lambda_0^2}
\label{eq:2.14} \\
\sigma(\mu_0,u) &=& -\frac{g^2}{4\pi^2}
\mu_0^2 \left( C_a (-\frac{u^2}{\mu_0^2}
+\ln\frac{u^2}{\mu_0^2}-\frac{5}{12})
+T_fN_f(\frac{1}{3}+\frac{m^2}{\mu_0^2}) \right)
\nonumber
\,,\end{eqnarray}
where the renormalization condition, Eq.~(\ref{eq:2.5}), 
at scale $\lambda_0$ is satisfied. 
Recall, this solution, Eq.~(\ref{eq:2.14}), 
describes an effective gluon mass at zero gluon momentum.
Since the effective Hamiltonian generated by flow equations 
is defined at cutoff scale $\lambda\rightarrow 0$,
the resulting gluon mass equals the 'physical' mass,  
$\mu^2_{ren}(\lambda=\lambda_0=0)=\mu_0^2$. In particular,
when the 'physical' mass is set to zero, $\mu_0=0$,
the effective QCD Hamiltonian has zero mass gauge fields, 
and our unitary transformation becomes gauge invariant.
Then the Kinoshita-Lee-Nauenberg's theorem
\cite{KinoshitaLeeNauenberg} is applicable, 
according to which IR singularities
associated with soft collinear gluons are cancelled exactly 
only for color singlet, gauge invariant matrix elements.
This was demonstrated in detail perturbatively for $e^+e^-$ annihilation
amplitude \cite{authors}. 
In our situation this zero gluon mass, gauge invariant limit 
also produces IR divergent quark and gluon self-energies. 
This in turn generates color confinement because 
the infinite self-energies suppress the quark and gluon propagation 
amplitudes. However, and, consistent with the above theorem, 
in the Bethe-Salpeter equation (and the simpler Tamm-Dancoff equation) 
the IR divergent contributions from the kinetic self-energy and potential
interaction kernel parts cancel exactly for color singlet states.
Our generated two-body $q\bar{q}$ and $gg$ potentials 
satisfy these conditions providing finite masses for singlet states
of mesons and glueballs. 
   
From Eq.~(\ref{eq:2.14}), 
one has in the limit $\mu_0\rightarrow 0$
\begin{eqnarray}
\sigma=\lim_{\mu_0\to 0}\sigma(\mu_0,u)=u^2\frac{g^2C_a}{2\pi^2}
\,,\label{eq:2.15}\end{eqnarray}
where, as shown below, 
$\sigma$ plays the role of the string tension 
between quark and antiquark. 
In Eq.~(\ref{eq:2.14}), $u^2$ determines 
the rate at which the effective gluon mass
asymptotically approaches  
the 'physical' value, $\sim u^2\ln(\lambda/\lambda_0)$.

Though this paper addresses confinement,
we are not yet able to suggest a physical picture
for how confinement emerges in the light-front frame.
However, we note that QCD has a non-abelian gauge group 
and strong interactions, which distinguishes it from QED.
Both canonical light-front 
QCD and QED Hamiltonians have the instantaneous 
interaction, $1/x^2$, which has a singular behavior for small $x$. 
For QED a small $x$ regularization   
scheme can be found where divergent contributions from this instantaneous
term are canceled in the matrix elements \cite{ZhangHarindranath}. 
This is not true for QCD \cite{ZhangHarindranath}. 
Due to the non-abelian triple-gluon vertex 
it is necessary to introduce a regulating scale $u$
for the divergences at $x\sim 0$ and $x\sim 1$ 
in the gluon self-energy (the quark loop is regulated in the same fashion
in order to combine quark and gluon loops). 
In fact, the string tension in Eq.~(\ref{eq:2.15})
is proportional to the Casimir operator in the adjoint representation
reflecting this non-abelian character.   
There is no such scale in the QED self-energy operators, which
are all regulated by a current electron mass $m$. 
One can therefore argue that the scale $u$ carries information
about strongly interacting fields and is present only in QCD.

In asymptotic free theories, such as QCD, the regulating scale 
can be related to the gauge invariant scale 
using the Callan-Symanzik equation.
In our case, the scale $u$ can be expressed 
in terms of $\Lambda_{QCD}$ by
solving the third order flow equations for 
the strong coupling constant, $\alpha_s(\lambda)$.
Such calculations have been recently performed 
for an asymptotic free toy model \cite{Glazek2} and
for the three-gluon vertex in QCD \cite{Glazek3}.
However, here we forgo this procedure and simply assume
$u$ is given by the hadron scale, $u\sim \Lambda_{hadron}$, 
that also establishes the string tension scale. 

The asymptotic behavior of the effective gluon mass 
near the renormalization point   
is important for selfconsistently
solving the flow equations for effective interactions, 
Eq.~(\ref{eq:1.6}), at vanishing gluon momenta. 
In the next subsection, this dependence (Eq.~(\ref{eq:2.14})) 
is used to find an effective 
quark potential generated by the flow equations.

\subsection{Effective quark-antiquark interaction}

Eliminating the quark-gluon coupling 
generates an effective interaction 
in the quark-antiquark sector 
(the second equation in Eq.~(\ref{eq:1.6})).
The energy transfers (i.e. energy denominators) 
along the quark $p_1\rightarrow p_2$
and antiquark $p_2^{\prime}\rightarrow p_1^{\prime}$ lines
are given by 
$D_1=p_1^{-}-p_2^{-}-(p_1-p_2)^{-}$ and 
$D_2=p_2^{\prime -}-p_1^{\prime -}
-(p_2^{\prime}-p_1^{\prime})^{-}$, 
respectively.
We denote the exchange momentum as $q=(q^+,q_{\perp})=p_1-p_2$.
These energy transfers can be related to the corresponding square 
of the four-momenta as $Q_1^2=(p_1-p_2)^2=-q^+D_1$ and 
$Q_2^2=(p_2^{\prime}-p_1^{\prime})^2=-q^+D_2$
\cite{Gubankova}.
In the light-front formulation, the scalar product is defined
$p\cdot k=(p^+k^- +p^-k^+)/2-p_{\perp}\cdot k_{\perp}$
and the four-momentum transfers for
quark momentum $(x,k_{\perp})$ 
and antiquark momentum $(1-x,-k_{\perp})$ are given by
\begin{eqnarray}
Q_1^2(\lambda) &=& Q_1^2+\mu^2_{ren}(\lambda)
\nonumber\\
Q_2^2(\lambda) &=& Q_2^2+\mu^2_{ren}(\lambda)
\,,\label{eq:2.16}\end{eqnarray} 
where the asymptotic form for 
the renormalized effective gluon mass, 
$\mu_{ren}^2(\lambda)=\mu_0^2+
\sigma(\mu_0,u)\ln(\lambda^2/\lambda_0^2)$
(Eq.~(\ref{eq:2.14})), enters. 
The four-momentum transfers are now
dependent on the cutoff $\lambda$ and for 
zero gluon mass are given \cite{Gubankova} by
\begin{eqnarray}
Q_1^2 &=& \frac{(x'k_{\perp}-xk'_{\perp})^2+m^2(x-x')^2}{xx'}
\nonumber\\
Q_2^2 &=& Q_1^2|_{x\rightarrow (1-x);~x'\rightarrow (1-x')}
\,.\label{eq:2.17}\end{eqnarray} 
As shown below, in the instant formulation, 
energy transfers with zero gluon mass, 
Eq.~(\ref{eq:2.17}), describe a three-momentum exchange, 
$\vec{q}^{~2}$ with $\vec{q}(q_z,q_{\perp})$, 
while energy transfers with non zero cutoffs, 
Eq.~(\ref{eq:2.16}), correspond
to an effective energy (with effective mass) exchange,
$\vec{q}^{~2}+\mu_{ren}^2(\lambda)$.
Near the renormalization point the energy transfers 
given by Eq.~(\ref{eq:2.16}) and Eq.~(\ref{eq:2.17}) 
coincide asymptotically, i.e. 
for $\lambda\to\lambda_0\sim 0$ and $\mu_0\sim 0$,
$Q_i^2(\lambda)\to Q_i^2$, $i=1,2$.
Using Eq.~(\ref{eq:2.16}), the gluon four-momentum
can be written as
\begin{eqnarray} 
q_{\mu}=p_{1\mu}-p_{2\mu}+\frac{\eta_{\mu}}{2q^+}Q_1^2(\lambda)
=p'_{2\mu}-p'_{1\mu}+\frac{\eta_{\mu}}{2q^+}Q_2^2(\lambda)
\,,\label{eq:2.23a}\end{eqnarray} 
where the light-front unit vector $\eta_{\mu}$
is $\eta^{\mu}=(2^{-},0^{+},0_{\perp})$,
with $\eta\cdot k=k^+$.
In subsequent calculations, we use
the average and difference of four-momenta transfers 
\begin{eqnarray}
Q^2(\lambda) &=& (Q_1^2(\lambda)+Q_2^2(\lambda))/2
= Q^2+\mu_{ren}^2(\lambda)
\nonumber\\
\delta Q^2(\lambda) &=& (Q_1^2(\lambda)-Q_2^2(\lambda))/2
=\delta Q^2
\,,\label{eq:2.24}\end{eqnarray}
where $Q^2=(Q_1^2+Q_2^2)/2$ and $\delta Q^2=(Q_1^2-Q_2^2)/2$.   

Solving the flow equations selfconsistently, 
the energy transfers for non-zero $\lambda$ define
effective interactions between quarks 
as specified by Eq.~(\ref{eq:1.6}).   
Thus, the similarity formfactors in each vertex 
$f(Q_1^2(\lambda);\lambda)$ and 
$f(Q_2^2(\lambda);\lambda)$
provide an effective interaction (see below).
However, for large $\lambda$ only high energies 
are eliminated and the momenta transfer
cutoff dependence, $Q_i^2(\lambda)$, is minimal.
Reducing $\lambda$, energy transfers are eliminated 
continuously from high to low energy.
Only for small $\lambda$, and 
corresponding momentum transfers, is the
asymptotic behavior of $Q_i^2(\lambda)$ essential. 
This is important in solving 
for the effective interaction, since
(see subsection $2.2$)
this asymptotic behavior (the cutoff dependent gluon 
effective energy at small $\lambda$)
eliminates the quark-gluon coupling
even for vanishing gluon momenta, i.e.
when $Q_1^2=Q_2^2=0$ ($x=x'$ and $k_{\perp}=k'_{\perp}$). 
If $Q^2=0$ and $\mu_0\rightarrow 0$, 
the similarity formfactor 
$f={\rm exp}(-Q^2(\lambda)/\lambda^2)$, 
with $Q^2(\lambda)$ given by Eq.~(\ref{eq:2.16}),
decays at small $\lambda\rightarrow 0$ 
(with $\lambda> \lambda_0$) as
\begin{eqnarray}
f\sim \left(\frac{\lambda^2}{\lambda_0^2}\right)^{
(-\sigma/\lambda^2)}
\,.\label{eq:2.18}\end{eqnarray}
Note this is a power suppression, not exponential,
with $\sigma$ specified by Eq.~(\ref{eq:2.15}).
Only for $\lambda=\lambda_0$ does $f=1$,
however, for other values, $\lambda\neq\lambda_0$,
$f$ decays and 
the integral over $\lambda$ in Eq.~(\ref{eq:1.6})
is finite in the effective interaction. 

In the light-front formulation, an effective quark-antiquark 
interaction is given by a matrix element of an effective 
light-front QCD Hamiltonian in the $q\bar{q}$-sector ($Q$-space) 
and includes dynamical interactions generated by flow equations
(the second equation in Eq.~(\ref{eq:1.6})) as well as 
instantaneous terms present in the 
the light-front gauge. These terms are emboddied
in the effective $q\bar{q}$-interaction   
\begin{eqnarray}
V_{q\bar{q}}=-4\pi\alpha_s C_f 
\langle\gamma^{\mu}\gamma^{\mu'}\rangle
\lim_{(\mu_0,\lambda_0)\rightarrow 0} B_{\mu\mu'}
\,,\label{eq:2.19}\end{eqnarray}
where $C_f=T^aT^a=(N_c^2-1)/2N_c$ is Casimir operator 
in the fundamental representation, 
and the current-current term in the exchange channel
is given by
(quark helicity notation is suppressed)
\begin{eqnarray}
\langle\gamma^{\mu}\gamma^{\mu'}\rangle
=\frac{\left( \bar{u}(-k_{\perp},(1-x))\gamma^{\mu}
u(k_{\perp},x)\right)
\left(\bar{v}(k'_{\perp},x')\gamma^{\mu'}
v(-k'_{\perp},(1-x'))\right)}
{\sqrt{xx'(1-x)(1-x')}}
\,.\label{eq:2.20}\end{eqnarray}
The interaction kernel, $B_{\mu\mu'}$, is 
analogous to the effective electron-positron interaction
in the light-front QED \cite{Gubankova},
except for the cutoff dependence in the energy transfers
\begin{eqnarray}
B_{\mu\mu'}=g_{\mu\mu'}\left(I_1+I_2\right)
+\eta_{\mu}\eta_{\mu'} \frac{\delta Q^2}{q^{+2}}
\left(I_1-I_2\right)
\,.\label{eq:2.21}\end{eqnarray}
Here $g_{\mu\mu'}$ is the light-front metric tensor 
and $\eta_{\mu}$ is given previously (see Eq.~(\ref{eq:2.23a})).
The cutoff dependent energy (four-momentum) transfer,
Eq.~(\ref{eq:2.16}),
along the quark and antiquark lines is included
in the integral 
\begin{eqnarray}
I_1=\int_0^{\infty}d\lambda\frac{1}{Q_1^2(\lambda)}
\frac{df(Q_1^2(\lambda);\lambda)}{d\lambda}
f(Q_2^2(\lambda);\lambda)
\,.\label{eq:2.22}\end{eqnarray}
$I_2$ is obtained by interchanging indices $1$ and $2$ 
in Eq.~(\ref{eq:2.22}).   
As discussed previously, 
the sensitivity to the renormalization scheme 
and renormalization point is eliminated by setting 
both $\mu_0$ and $\lambda_0$ to zero after integration.

For the following three explicit forms for the similarity function,
the leading behavior of the integral factor, Eq.~(\ref{eq:2.22}), 
is given by
\begin{eqnarray} 
&& (1)~ Exponetial  
\nonumber \\
 f &=&
{\rm exp}\left( -Q^2(\lambda)/\lambda^2\right),
\hspace{1cm} 
I_1=\frac{1}{Q_1^2(\lambda_0)+Q_2^2(\lambda_0)}
\left(1+\frac{\sigma(\mu_0,u)}{Q_1^2(\lambda_0)}
\right) 
\nonumber\\
&& (2)~ Gaussian 
\nonumber \\
 f &=&
{\rm exp}\left(-Q^4(\lambda)/\lambda^4\right),
\hspace{1cm} 
I_1=\frac{Q_1^2(\lambda_0)}{Q_1^4(\lambda_0)+Q_2^4(\lambda_0)}
\left(1+\frac{\sigma(\mu_0,u)}{Q_1^2(\lambda_0)}
\right) 
\nonumber\\
&& (3)~ Sharp 
\nonumber \\
f &=&
\theta(\lambda^2-Q^2(\lambda)),
\hspace{1cm} 
I_1=\frac{\theta(Q_1^2(\lambda_0)-Q_2^2(\lambda_0))}
{Q_1^2(\lambda_0)}
\left(1+\frac{\sigma(\mu_0,u)}{Q_1^2(\lambda_0)}
\right) 
\,,\label{eq:2.23}\end{eqnarray}
where from Eq.~(\ref{eq:2.16})
the four-momenta are given by 
$Q_i^2(\lambda_0) = Q_i^2+\mu_0^2$ with $i=1,2$.
Several properties of the similarity function were used
to approximate the integral in Eq.~(\ref{eq:2.22}).  
The similarity function $f(z)$ decays
for arguments $z\geq 1$, thus the integral saturates
for values 
$0\leq\lambda^2 \leq (Q_1^2(\lambda_0),Q_2^2(\lambda_0))$
at small energy transfer. 
Also, $\mu_0\sim \lambda_0 \sim 0$. 
Note Eq.~(\ref{eq:2.23})
has the form consistent with the exchange of a 
gluon with an effective mass parameter $\mu_0$
between quark and antiquark at distances 
$r\sim 1/\mu_0$.
The resulting $q\bar{q}$-interaction is defined in the limit
of vanishing mass parameter $\mu_0\rightarrow 0$
and at the zero renormalization cutoff point $\lambda_0\rightarrow 0$, 
Eq.~(\ref{eq:2.19}).
In this limit, the average four-momentum transfer 
reduces to $\lim_{(\mu_0,\lambda_0) 
\rightarrow 0} Q^2(\lambda_0)=Q^2$, Eq.~(\ref{eq:2.24}),
and the string tension is given by
$\lim_{(\mu_0,\lambda_0) \rightarrow 0}\sigma(\mu_0,u)=\sigma$,
Eq.~(\ref{eq:2.15}). 
Using Eqs.~(\ref{eq:2.21}) and (\ref{eq:2.23}),
the resulting interaction kernel 
for the three similarity function choices is
\begin{eqnarray}
&& (1)~ Exponential
\nonumber \\
\lim_{(\mu_0,\lambda_0)\rightarrow 0} B_{\mu\mu'} &=& 
g_{\mu\mu'}\left(
\frac{1}{Q^2}
+\frac{\sigma}{Q^4}\right)
+\left(\frac{g_{\mu\mu'}}{Q^2}
-\frac{\eta_{\mu}\eta_{\mu'}}{q^{+2}}\right)
\frac{\sigma}{Q^2}
\frac{\delta Q^4}{Q^4-\delta Q^4} 
\nonumber\\
&& (2)~ Gaussian
\nonumber \\
\lim_{(\mu_0,\lambda_0)\rightarrow 0} B_{\mu\mu'} &=& 
g_{\mu\mu'}\left(
\frac{1}{Q^2}
+\frac{\sigma}{Q^4}\right)
-\left(\frac{g_{\mu\mu'}}{Q^2}
(1+\frac{\sigma}{Q^2})
-\frac{\eta_{\mu}\eta_{\mu'}}{q^{+2}}\right)
\frac{\delta Q^4}{Q^4+\delta Q^4} 
\nonumber\\
&& (3)~ Sharp
\nonumber \\
\lim_{(\mu_0,\lambda_0)\rightarrow 0} B_{\mu\mu'} &=& 
g_{\mu\mu'}\left(
\frac{1}{Q^2}
+\frac{\sigma}{Q^4}\right)
-\left(
\frac{g_{\mu\mu'}}{Q^2}
(1+\frac{\sigma}{Q^2}
(1+\frac{Q^2}
{Q^2+\left|\delta Q^2\right|}))\right.
\nonumber\\
&-& \left.\frac{\eta_{\mu}\eta_{\mu'}}{q^{+2}}
(1+\frac{\sigma}{Q^2}
\frac{Q^2}
{Q^2+\left|\delta Q^2\right|})
\right)
\frac{\left|\delta Q^2\right|}
{Q^2+\left|\delta Q^2\right|} 
\,,\label{eq:2.25}\end{eqnarray}
where $Q^4=(Q^2)^2$ and $\delta Q^4=(\delta Q^2)^2$.
The four-momenta in Eq.~(\ref{eq:2.25})
can be represented 
in the 'mixed' light-front and instant 
representations as 
\begin{eqnarray}
Q^2 &=& \vec{q}^{\, 2}
-\frac{1}{4}(2x-1)(2x'-1) (M_1-M_2)^2
\nonumber\\
\delta Q^2 &=& -\frac{1}{2}(x-x') (M_1^2-M_2^2)
\,,\label{eq:2.28}\end{eqnarray}
where, together with the light-front momentum 
$(x,k_{\perp})$, the instant momentum $\vec{k}=(k_z,k_{\perp})$
enters with the connection
\begin{eqnarray}
x=\frac{1}{2}\left(1+\frac{k_z}{\sqrt{\vec{k}^{\, 2}+m^2}} \right)
\,,\label{eq:2.27}\end{eqnarray}
and $\vec{k}^{\, 2}= k_{\perp}^2+k_z^2$.
The gluon three-momentum transfer is given by
$\vec{q}=\vec{k}-\vec{k}^{\,\prime}=(q_z,q_{\perp})$   
and the total energies of the inital and final 
$q\bar{q}$-states are given by 
\begin{eqnarray}
M_1^2 &=& 4(\vec{k}^{\, 2}+m^2) 
\nonumber\\
M_2^2 &=& 4(\vec{k}^{\,\prime 2}+m^2)  
\,.\label{eq:2.29}\end{eqnarray}

Our results, Eq.~(\ref{eq:2.25}), are not rotationally
invariant due to truncation of Fock-space.
However, on the energy shell 
($M_1^2=M_2^2$ and $\delta Q^2=0$), the second term in 
Eq. (\ref{eq:2.25}) is identically zero and 
the effective $q\bar{q}$ interaction is given by
the first term
\begin{eqnarray}
V_{q\bar{q}}=-\langle\gamma^{\mu}\gamma_{\mu}\rangle
\left(C_f\alpha_s\frac{4\pi}{\vec{q}^{\, 2}}
+\sigma_f\frac{8\pi}{\vec{q}^{\, 4}}
\right) 
\,.\label{eq:2.26}\end{eqnarray}
It is significant to note that all three choices yield
the same effective interaction.
Here $\sigma_f=\sigma\alpha_sC_f/2$
with $\sigma$ given by Eq.~(\ref{eq:2.15}).
Further, and interestingly, we also obtain a Cornell type
interaction with Coulomb and linear confining
potentials 
\begin{eqnarray}
V_{q\bar{q}}=\langle\gamma^{\mu}\gamma_{\mu}\rangle
\left(-C_f\frac{\alpha_s}{r}+\sigma_f\cdot r
\right)
\,.\label{eq:2.30qq}\end{eqnarray}
If the external particles are all on the energy shell, our light-front 
formulation yields a rotationally invariant potential as expected.
The confining term in Eq.~(\ref{eq:2.30qq}) 
(or Eq.~(\ref{eq:2.26})),
arises from elimination of the quark-gluon coupling
at small gluon momenta, which is governed by 
the asymptotic behavior 
of the effective gluon mass at small cutoff.
It is also significant to note that the same quark effective
interaction, Eqs.~(\ref{eq:2.26},\ref{eq:2.30qq}),
emerges in the limit of static (infinite heavy
$m\rightarrow \infty$) quarks.

The $N_c$ behavior of the confining term in Eq.~(\ref{eq:2.30qq})
with the string tension given by
\begin{eqnarray}
\sigma_f=\frac{u^2}{4\pi^2}\frac{g^4C_aC_f}{4\pi} =
\frac{u^2}{4\pi^2}\frac{g^4(N_c^2-1)}{8\pi}
\,,\label{eq:2.30Nc}\end{eqnarray}
can be compared with available $2+1$ lattice data \cite{Teper}  
for the expectation value of the Wilson loop in the fundamental representation,
$\langle W_{f}(C)\rangle =\exp(-\sigma_{2+1}S_C)$,
where $S_C$ is the area of the loop $C$. 
Though the data are for $2+1$
the same $N_c$ behavior is expected for $3+1$ and higher dimensions \cite{Teper}.
Monte Carlo calculations of the string tension in $2+1$ give
the values $\sqrt{\sigma_{2+1}}/g^2 = 0.335,~ 0.553,~ 0.758,~0.966$
for the gauge groups $SU(2),~SU(3),~SU(4)$ and $SU(5)$ respectively \cite{Teper}
(note that the coupling constant in $2+1$ has dimension, $g^2\sim$energy).
The corresponding values calculated from Eq.~(\ref{eq:2.30Nc})
are $2\pi\sqrt{\sigma_f}/(ug^2) = 0.345,~ 0.564,~ 0.772,~ 0.977$. We see
that there is excellent agreement (up to $\sim 3~\%$) for this $N_c$ behavior 
between Eq.~(\ref{eq:2.30Nc}) and the Monte Carlo results.
It is further interesting to notice that our analytic expression 
for the string tension, Eq.~(\ref{eq:2.30Nc}), has the appropriate 
$N_c$ dependence as expected from large $N_c$ calculations.

Similarly the quark-quark potential can be obtained 
and is related to the quark-antiquark potential by
\begin{eqnarray}
V_{qq}=-\frac{1}{2}V_{q\bar{q}}
\,,\label{eq:2.30qq2}\end{eqnarray}
since only the commutator $[b^{\dagger}ba,b^{\dagger}ba^{\dagger}]$
contributes to $V_{qq}$, while $[b^{\dagger}ba,d^{\dagger}da^{\dagger}]$
and $[d^{\dagger}da,b^{\dagger}ba^{\dagger}]$ contribute to $V_{q\bar{q}}$.
Assuming the relation Eq.~(\ref{eq:2.30qq2}), Basdevant and Boukraa
\cite{BasdevantBoukraa} showed that the ground state of baryons can be calculated
with good accuracy.
   
Note, because we use perturbation theory, 
we do not claim to have completed the derivation
of a confining potential in QCD.
However, our $1/\vec{q}^{~4}$ quark-antiquark potential
might be a precursor to quark confinement.
Perhaps equally as important, our results conflict with
the notion that a weak-coupling expansion will never produce 
a confining potential.  

Finally, there are corrections $O\left( \delta Q^2/Q^2 \right)$ 
to the leading effective interaction, which  
depend on the direction $\vec{q}$
approaches zero (as was investigated in \cite{Gubankova}). 
An important limiting case is the collinear limit
\begin{eqnarray}
q^+\rightarrow 0
\,.\label{eq:2.29a}\end{eqnarray}
This is special for light-front calculations
and may cause divergences.
From Eq.~(\ref{eq:2.28}) in this limit, 
$\delta Q^2\sim q^+$, and 
for sufficiently smooth similarity functions $f(z)$,
like exponential or Gaussian, the effective interaction
Eq.~(\ref{eq:2.25})
does not contain a collinear singularity, because
$\left(\eta_{\mu}\eta_{\mu'}/q^{+2}\right)\cdot\delta Q^4$ 
is finite.
Thus the interaction only becomes singular if $\vec{q}$ 
approaches zero as $1/\vec{q}^{\,2}$ 
('Coulomb' singularity)
or $1/\vec{q}^{\,4}$ ('confining' singularity), 
namely the leading behavior given by 
Eq.~(\ref{eq:2.26}). Note that for mass spectrums   
both singularities are controllable.  
The 'Coulomb' singularity is integrable, because
the integral $\int d^3q/q^2$ is finite for small $q$
and the 'confining' singularity can be regulated
in the IR.
However, this is not true
for the sharp cutoff where the $\eta_{\mu}\eta_{\mu'}$ term
diverges as $1/q^+$, because 
it appears in the combination
$\left(\eta_{\mu}\eta_{\mu'}/q^{+2}\right)\cdot|\delta Q^2|$.
We do not attach physical significance to this singularity,
which is a consequence of the arbitrary cutoff choice
leading to a singular unitary
transformation generator at small momenta Eq.~(\ref{eq:1.3}).
In summary, the collinear singularity 
is completely absent for a smooth cutoff 
and only the rotationally invariant 
part of the effective interaction 
Eq.~(\ref{eq:2.26}) remains
in the limit $\vec{q}\rightarrow 0$, 
where the dominant contribution to the spectrum is obtained.
Corrections from the $\eta_{\mu}\eta_{\mu'}$ term are finite
and shift the entire spectrum by a constant since
they are diagonal in spin space.

As numerically documented in the positronium
QED effective interaction application \cite{Gubankova},  
rotational symmetry holds with high accuracy 
even if the on-energy-shell condition for the external particles
is removed. This holds for smooth 
cutoff functions and even for a sharp cutoff if the collinear
singular part is subtracted.
Based on our analysis here, we may anticipate 
similar results for the QCD effective interaction.

\subsection{Effective gluon-gluon interaction} 

In full analogy with section $3.2$, eliminating the
three-gluon coupling generates an effective interaction
between two gluons. Again, in leading order, 
the asymptotic behavior of the effective mass 
of the exchanged gluon is included. Using 
the same notation as section $3.2$,
an effective gluon-gluon interaction
is again generated by the flow equations.
It contains both dynamical and instantaneous gluon exchange
and is given by
\begin{eqnarray}
V_{gg}=-4\pi\alpha_s C_a 
\langle\Gamma^{\mu}\Gamma^{\mu'}\rangle
\lim_{(\mu_0,\lambda_0)\rightarrow 0} B_{\mu\mu'}
\,,\label{eq:2.30}\end{eqnarray}
where now $C_a\delta_{ab}=f^{acd}f^{bcd}=N_c\delta_{ab}$ 
is the Casimir operator in the adjoint representation. 
Here, the current-current term in the exchange channel
is given by
\begin{eqnarray}
\langle\Gamma^{\mu}\Gamma^{\mu'}\rangle = \frac{
\Gamma^{\mu\nu\rho}(-q,p_1,-p_2)
\Gamma^{\mu'\nu'\rho'}(q,p'_1,-p'_2)}
{\sqrt{xx'(1-x)(1-x')}}
\epsilon_{\nu}(p_1)\epsilon^{*}_{\rho}(p_2)
\epsilon_{\nu'}(p'_1)\epsilon^{*}_{\rho'}(p'_2)
\,,\label{eq:2.31}\end{eqnarray}
where
\begin{eqnarray}
\Gamma^{\mu\nu\rho}(-q,p_1,-p_2)
\Gamma^{\mu'\nu'\rho'}(q,p'_1,-p'_2) &=&  
\left((p_1+p_2)^{\mu}g^{\rho\nu}+(-2p_1)^{\rho}g^{\mu\nu}       
+(-2p_2)^{\nu}g^{\mu\rho} \right)
\label{eq:2.32}\\
&\times&
\left((p'_1+p'_2)^{\mu'}g^{\rho'\nu'}
+(-2p'_1)^{\rho'}g^{\mu'\nu'}       
+(-2p'_2)^{\nu'}g^{\mu'\rho'} \right)
\nonumber
\,,\end{eqnarray}
and the light-front momenta are $p_1=(x,k_{\perp})$, 
$p_1^{\prime}=(1-x,-k_{\perp})$ and $p_2=(x',k'_{\perp})$,
$p'_2=(1-x',-k'_{\perp})$. 
The gluon polarizations are omitted 
for simplicity.
The interaction kernel $B_{\mu\mu'}$ 
is given by Eq.~(\ref{eq:2.21}), 
which is again defined
for different cut-off functions in Eq.~(\ref{eq:2.25}). 
The polarization vector property, 
$q\cdot\epsilon=\eta\cdot\epsilon=0$, 
and representation for the gluon momentum,  
Eq.~(\ref{eq:2.23a}), have been used
to simplify the above expressions.

Both quark-antiquark and gluon-gluon interactions have the same
kernel $B_{\mu\mu'}$, and differ only in prefactors.
Following arguments detailed in section $3.2$, 
the leading gluon-gluon interaction 
is also independent of cut-off function form
and is given by
\begin{eqnarray}
V_{gg}=\langle\Gamma^{\mu}\Gamma_{\mu}\rangle
\left(-C_a\frac{\alpha_s}{r}
+\sigma_a\cdot r
\right) 
\,,\label{eq:2.33gg}\end{eqnarray}
which is, again, a Coulomb plus linear potential
between two color sources in the adjoint representation
with $\sigma_a=\sigma\alpha_sC_a/2$.
The only difference between the $q\bar{q}$, Eq.~(\ref{eq:2.30qq}),  
and $gg$, Eq.~(\ref{eq:2.33gg}), potentials
is given by the ratio of the fundamental to adjoint 
Casimir operators
\begin{eqnarray}
V_{q\bar{q}}/V_{gg}=C_f/C_a=(N_c^2-1)/2N_c^2=4/9
\,,\label{eq:2.34}\end{eqnarray}
for $SU(3)$, since both potentials have the same Coulomb 
plus confining behavior.

\section{Concluding discussion}

An effective QCD Hamiltonian in the light-front gauge
has been obtained, solving the flow equations for the two lowest
Fock sectors selfconsistently. It has been shown that
it is possible to eliminate the minimal quark-gluon interaction
by using a continuous unitary transformation. In this elimination,
the coupling functions of the Hamiltonian described by the flow equations
are renormalized. 
In the two lowest Fock sectors this change 
of the couplings corresponds to the renormalization
of the one-particle energies and to the generation 
of effective interactions between gluons and quarks, in particular
the quark-antiquark interaction.  
In obtaining the flow differential equations,   
intermediate states with more than three particles 
were omitted. Fock number truncation 
in intermediate states is quite different from a 
perturbation coupling constant treatment and more similar to
the Tamm-Dancoff approach.

Our approach has several advantages. First,
the original gauge field coupling can be completely
eliminated, even when the states connected by this
interaction are degenerate. The continuous transformation 
is designed such that the transformed Hamiltonian does not
contain any interactions between one (anti)quark and the creation 
or annihilation of one gluon. These unwanted interactions,
connecting states with energy differences less than
a cutoff scale $|E_p-E_q|\leq\lambda$, are present
in the similarity renormalization approach 
because single particle energies are not renormalized. 
They can mix low and high Fock sectors,
and are not amendable to a perturbative treatment. 
Further, ignoring these low-energy interactions 
may break gauge invariance, 
and, in the light-front formulation,  
rotational symmetry as well.
Second, an effective quark-antiquark interaction 
is rotationally symmetric when the external particles are
on the energy shell. At small gluon momenta $q$, also 
the collinear singular terms $\sim 1/q^+$ 
and $\sim 1/q^{+\, 2}$ cancel. Most interestingly,
in addition to a Coulomb term $1/q^2$,
which can be obtained in second order,
there is also a more singular, confining term $1/q^4$.  
Our induced $q\bar{q}$-interaction also differs  
from the similarity renormalization result, 
where the collinear singular
part of the remnant instantaneous interaction
($\sim 1/q^{+\, 2}$) yields a logarithmic potential
which is not rotationally symmetric.
These differences stem from the 
coupling's flow parameter dependence. 

Further, due to complete elimination of 
the quark-gluon coupling,
the flow equation for an effective quark-antiquark interaction
can be integrated for all cutoffs including $\lambda=0$.
In the similarity renormalization approach 
one removes couplings perturbatively
reaching a minimum scale (cut-off),
below which perturbation theory
breaks down. The value of this cutoff depends on the 
problem considered, and might be ambiguous. 
For QCD this cutoff 
introduces a scale in the theory, 
which breaks gauge 
and rotational invariance \cite{BrisudovaPerry}. 
In our approach, the regulator of small light-front $x$
establishes a scale in the effective theory
corresponding to the string tension 
in the effective quark-antiquark potential.
Besides the nonzero scale, 
the resulting renormalized gluon mass
vanishes asymptotically, maintaining gauge invariance.
Also, the effective quark-antiquark interaction
is rotationally symmetric when the external particles are on the
energy shell.
A shortcomming in this work is that 
the small light-front $x$ cutoff scale $u$
enters as an input parameter, fitted to the string tension
from lattice calculations. This can be improved by  
relating the cutoff $u$ to the renormalization group invariant
scale $\Lambda_{QCD}$ which necessitates
higher Fock sector intermediate states.
This would also permit confirmation of our regularization scheme 
independent results.

Our ultimate goal is to solve the coupled chain
of flow equations in different sectors selfconsistently.
As shown in this work, even an approximate solution
of the gluon gap equation combined with the flow equation
for the effective $q\bar{q}$ interaction 
provides new information beyond standard perturbation theory.
The next step is to address the quark sector,
formulating the quark gap equation and 
obtaining the renormalized light-front quark mass 
and attending improved effective quark interaction.

Finally, reflecting the above discussions,
we note that the light-front flow equations avail
themselves to selfconsistent computer generated solution.
It would be quite interesting to confront hadronic data
with large scale, but feasible applications 
of this approach.

In summary, the light-front formulation appears quite 
viable for examining the long range aspects of QCD.
Implementing the flow equations within this framework has
generated new insight regarding nonperturbative phenomena 
including confinement. 

\section*{Acknowledgments}
This work was supported by U.S. DOE grants  
DE-FG02-97ER41048 and DE-FG02-96ER40947.

\end{document}